# An Overview of Modern Machine Learning Methods for Effect Measure Modification Analyses in High-Dimensional Settings.

Michael Cheung, Anna Dimitrova, Tarik Benmarhnia

# Abstract

A primary concern of public health researchers involves identifying and quantifying heterogeneous exposure effects across population subgroups. Understanding the magnitude and direction of these effects on a given scale provides researchers the ability to recommend policy prescriptions and assess the external validity of findings. Furthermore, increasing popularity in fields such as precision medicine that rely on accurate estimation of high-dimensional interaction effects has highlighted the importance of understanding effect modification. Traditional methods for effect measure modification analyses include parametric regression modeling with either stratified analyses and corresponding heterogeneity tests or including an interaction term in a multivariable model. However, these methods require manual model specification and are often impractical or not feasible to conduct by hand in high-dimensional settings. Recent developments in machine learning aim to solve this issue by automating heterogeneous subgroup identification and effect estimation. In this paper, we summarize and provide the intuition behind modern machine learning methods for effect measure modification analyses to serve as a reference for public health researchers. We discuss their implementation in R, provide annotated syntax and review available supplemental analysis tools by assessing the heterogeneous effects of drought on stunting among children in the Demographic and Health Survey data set as a case study.

# 1 Introduction

Effect measure modification (EMM) (or treatment effect heterogeneity) is present when there are differences in an exposure-outcome relationship across subgroups in a population and constitutes an important consideration for public health researchers (Vanderweele 2009). Said differently, we say that M is a modifier of the effect of X on Y when the average treatment effect of X on Y varies across levels of M. Since the average treatment effect of X on Y can be measured using various effect measures on either multiplicative or additive scales (e.g., risk difference, risk ratio), the presence of effect modification depends on the effect measure being used.

Understanding the effect of an exposure on a given outcome within population subgroups is important for two main reasons. First, it can guide intervention prioritization for those who will benefit more (depending on the scale of interest) from the treatment (VanderWeele and Knol 2014). For example, an EMM analysis may uncover that a particular subgroup in a study population stands most to benefit from a vaccine, providing policy makers reason to prioritize their vaccine uptake. This proves even more important in instances when resources are limited. EMM analyses can also determine if an exposure is harmful or beneficial to a subgroup when the population level effect is non-existent or trends in the opposite direction (VanderWeele and Knol 2014). Results of this type can reveal that current policy prescriptions are suboptimal or negatively impacting certain subgroups. Furthermore, discovery of effect modification advances the understanding of a potentially complex relationship between an exposure and outcome. In a medical practice setting, this can give professionals the ability to select a treatment that maximizes the probability of desired outcomes given the characteristics of a patient (National Research Council 2011). Second, quantifying EMM in a given exposure-outcome relationship is critical to external validity applications including transportability and generalizability analyses (Lesko et al. 2017). Indeed, the main reason for which an effect estimates in a given study population may not be generalizable (to the target population) or transportable to another population is because of a differential distribution of effect modifiers. Under the potential outcomes framework, the distribution of effect modifiers also has implications for the potential violation of the consistency assumption as if the distribution of effect modifiers differs across different versions of the treatment, the inference may not be generalizable to the target population (Rehkopf, Glymour, and Osypuk 2016).

Traditional methods for EMM analyses include two approaches: i) including an interaction term in a multivariable model; ii) conducting stratified analyses coupled with a heterogeneity test. The first traditional approach involves parametric regression modeling in which heterogeneity is assessed by an interaction term between the exposure variable and effect modifier(s) (VanderWeele and Knol 2014). The second traditional approach includes stratified analyses where separate models are run on subgroups to compare subgroup treatment effects using tests such as Wald or Cochran Q tests (Kaufman and MacLehose 2013). Under the potential outcomes' framework, the concepts of effect measure modification and interaction fundamentally differ. Indeed, when mobilizing the concept of interaction, we aim at manipulating both the exposure of interest X and a third variable of interest M hypothesizing a joint intervention, which requires that the identification assumptions (e.g. exchangeability, positivity and consistency) hold for both X and M, which is not the case when mobilizing the concept of effect modification. That being said, estimation techniques are similar for both concepts and all analytical details discussed in the following sections can theoretically apply to either effect measure modification or interaction analyses. However, several problems exist with these traditional methods.

These traditional approaches require manual specification of the effect modifiers and confounders, which is often burdensome or simply not feasible for high-dimensional and non-linear relationships. Moreover, the problem of multiple comparisons arises when the number of conducted analyses increases (Greenland 2008). Distributional assumptions may not hold with real world data, introducing bias and leading to incorrect conclusions about the estimated effects. These challenges prompt the use case for machine learning (ML) approaches that remove the requirement of manual specification and provide identification and estimation of EMM in a data-driven manner.

In the past decade, many ML methods have been proposed to address this need. Nonparametric tree-based methods, in both frequentist and bayesian frameworks, are some of the most developed and widely used approaches (Su et al. 2012; Athey and Imbens 2016; Powers et al. 2018; Wager and Athey 2018; Athey, Tibshirani, and Wager 2019; Chipman, George, and McCulloch 2010; J. L. Hill 2011; Hahn, Murray, and Carvalho 2020). Other methods traditionally used for prediction such as LASSO (Imai and Ratkovic 2013; Belloni, Chernozhukov, and Hansen 2013; Zhao, Small, and Ertefaie 2021) and neural networks (Shalit, Johansson, and Sontag 2017; Syrgkanis et al. 2019) have been adapted for EMM analyses but have been less used. Ensemble methods that combine multiple base algorithms like metalearners (Kunzel et al. 2019) and stacking (Nie and Wager 2020) have also been applied, and more recently designed for identifying and estimating heterogeneity (Lee, Bargagli-Stoffi, and Dominici 2021). These approaches make use of a variety of estimation, inference, and analytical techniques to assess EMM. As a result, a few studies have described and compared these methods to assess their performance and utility (Wendling et al. 2018; Dorie et al. 2019; McConnell and Lindner 2019; L. Hu et al. 2021; Liu 2022; A. Hu 2023). However, ML for EMM is a rapidly evolving field of study and there is a continuous need for interpretation of these methods, as well as guidance on applying them to real data. While a limited number of epidemiological studies used such approaches in the past few years, we are not aware of an up-to-date summary of some of the most commonly used methods as well as a guide about their implementation using an illustrative case study.

In this paper, we summarize and provide the intuition behind modern ML approaches for assessing high-dimensional effect modification. These include generalized random forests, bayesian additive regression trees, and bayesian causal forests. We do not consider ML methods that are used in the EMM literature, but are primarily concerned with prediction such as metalearners (e.g. S/T/X/R) and neural networks. While not an exhaustive list, these methods are the most widely used ML methods for EMM analyses at the time of this review. We discuss their implementation and corresponding available analysis tools in R (R Core Team 2023) with reproducible code using as a case study the effect of droughts on stunting in Sub-Sahara, relying on the Demographic and Health Survey (DHS) data set.

Lastly, we compare results across methods and to those of traditional methods. This overview aims to provide public health researchers a useful, straightforward resource for understanding and implementing ML methods for low or high-dimensional EMM analyses. Section 2 covers notation and the summary of each method. In section 3, we review the available tools and techniques for implementation of these methods, apply them to the DHS data, and compare results. Section 4 concludes with a discussion.

## 2 EMM ML Method Overview

We first briefly introduce the notation that will be used throughout this overview via Rubin's potential outcomes framework (Rubin 1974). Let $Y_i(Z_i)$ be the potential outcome under the exposure $Z_i = 0,1$ and $X_i$ be the covariate values for unit $i = 1, \ldots, n$. The effect of an exposure on a given outcome across a population is measured by the average treatment effect (ATE). To measure heterogeneous effects, the ATE is estimated conditioned on covariates that constitute population subgroups. This quantity is called this conditional average treatment effect (CATE). The ATE is then defined as $\mathbb{E}[Y_i(1) - Y_i(0)]$ and the CATE is defined as $\mathbb{E}[Y_i(1) - Y_i(0)|X_i = x]$ for covariates $x \in \mathbb{R}^p$. Because the CATE can be defined at different levels of granularity for $x$, we distinguish between a low level of granularity and a high level, that we refer to as an individual treatment effect (ITE). The ITE is defined identically to the CATE, with the exception that the dimension of the covariate space $p$ is large, say $q < p \leq P$ for some large $q$ where $P$ is the maximum dimension of the covariate space. We then define the dimension of the covariate space of the CATE as a value small enough that the corresponding subgroup is interpretable, i.e. $0 < p \leq q$. This distinction is important because the output of many EMM ML approaches is the ITE. As a consequence, the effects are not practical for the purpose of identifying effect modifiers. Lastly, throughout this review, we assume the standard causal assumptions of consistency ($Y_i = Z_i Y_i(1) + (1 - Z_i) Y_i(0)$), positivity ($P(Z_i = z|X_i = x) > 0$ for all $x$ with $P(X_i = x) \neq 0$), and unconfoundedness ($Y_i(1), Y_i(0) \perp Z_i|X_i$) (Hernán 2012) to make causal claims about observed effects. While the average treatment effects on the treated (ATT) $\mathbb{E}[Y_i(1) - Y_i(0)|Z_i = 1]$ is often the parameter of interest to researchers, we work with the ATE because the ATE = ATT under these assumptions (A. Wang, Nianogo, and Arah 2017). Table 1 summarizes this notation.

Table 1: *Glossary of terms*

| Estimand | Acronym | Description |
| --- | --- | --- |
| Average Treatment Effect | ATE | The expected difference between the potential outcomes in which the population is treated and untreated, i.e. $\mathbb{E}[Y_i(1) - Y_i(0)]$ |
| Conditional Average Treatment Effect | CATE | The expected difference between the potential outcomes in which the population is treated and untreated conditioned on a small number of covariates x, i.e. $\mathbb{E}[Y_i(1) - Y_i(0)|X_i = x]$ where $x \in \mathbb{R}^p$ for $0 < p \leq q, p, q \in \mathbb{Z}$ |
| Individual Treatment Effect | ITE | The expected difference between the potential outcomes in which the population is treated and untreated conditioned on a large number of covariates x, i.e. $\mathbb{E}[Y_i(1) - Y_i(0)|X_i = x]$ where $x \in \mathbb{R}^n$ for $q < p \leq P, q, p, P \in \mathbb{Z}$ |

## 2.1 Generalized Random Forests

Generalized random forests (GRF) was proposed by Athey et al. (2019) as the most recent development in a series of nonparametric tree-based methods for EMM (Athey and Imbens 2016; Wager and Athey 2018). The motivation behind these methods is twofold: data-driven estimation of heterogeneity and valid inference with confidence intervals for high-dimensional effect modification. The first proposed method was honest causal trees (HCT) which modified the traditional classification and regression tree (CART) to estimate ITEs instead of predicting outcomes (Athey and Imbens 2016). HCT implemented an "honest estimation" technique akin to traditional machine learning training-test splits in which the provided sample is divided into a set for tree growth and a set for ITE estimation. This technique reduces potential overfitting and decreases bias (Athey and Imbens 2016). Causal forests (CF) (Wager and Athey 2018) was then developed as an extension of HCT, incorporating the key components of Breiman's Random Forest (RF) framework (2001): bootstrap aggregating (bagging) across an ensemble of trees and random sampling of covariates used to determine splits.

GRF was introduced as the next iteration in the series to better handle cases of strong confounding (Athey, Tibshirani, and Wager 2019). At a high level, the algorithm recursively partitions the data on covariate splits that maximize heterogeneity to estimate ITEs. CF (and HCT) behaves similarly but abides by traditional RF bagging whereby predictions for a given vector of covariates $x$ are averaged across trees to obtain a final outcome. While intuitive, averaging fails to reduce bias for noisy solutions (Athey, Tibshirani, and Wager 2019). To address this problem, GRF uses an adaptive kernel method that performs locally weighted optimization to estimate ITEs. Each observation in the sample data is assigned a weight that represents how often the covariate values $X_i$ falls into the same terminal node as the given covariate vector $x$ (see figure 1). These weights depict neighborhoods with similar observations to $x$ and contribute to more stable ITE estimates than averaging. The figure below illustrates the weighting process. Each square in the left column depicts a tree in a forest with internal boxes representing terminal node partitions. The given $x$ is shown by a red triangle and the square in the right column shows the final observations with node sizes relative to their weights.

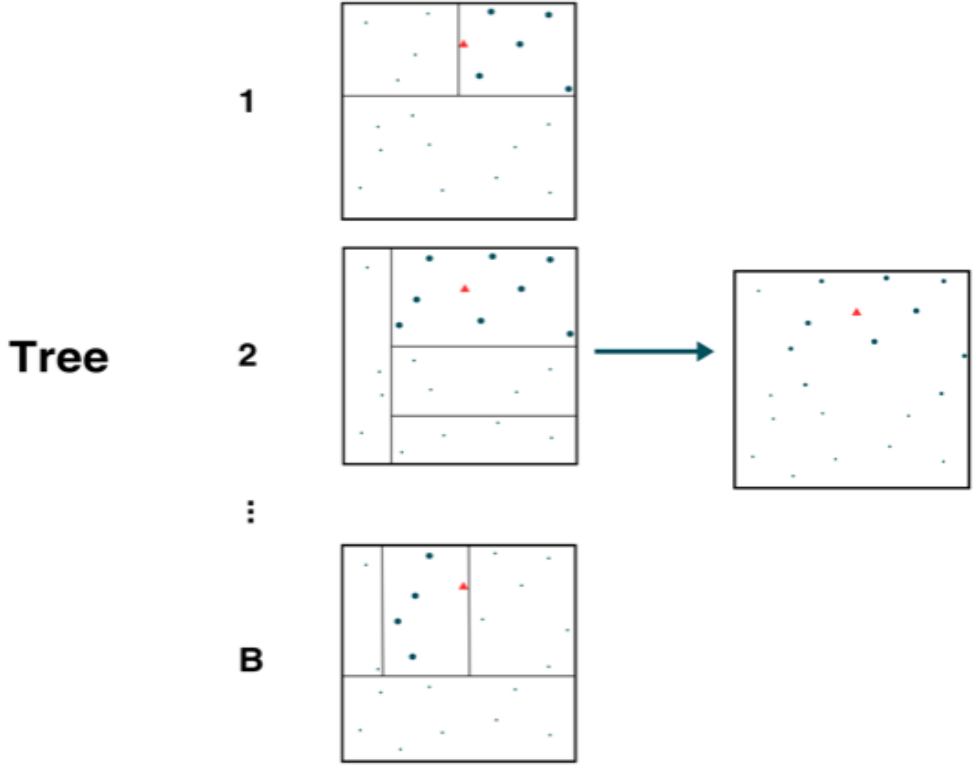

*Figure 1. Visualization of the Generalized Random Forest weighting procedure, inspired by fig. 1 from Athey et al. (2019)*

The algorithm can be summarized with the following steps:

1. Bootstrap the sample $B$ times.
2. For each $b$ in $B$, "honest split" $b$ into non-overlapping training and test sets.
3. For each training set, build a gradient tree $\mathcal{T}$. Trees are generated by maximizing the criterion $\Delta = \frac{n_{C_1} n_{C_2}}{n_P^2} \left( \hat{\theta}_{C_1} - \hat{\theta}_{C_2} \right)^2$ (with gradient-based approximations of $\hat{\theta}_{C_1}, \hat{\theta}_{C_2}$ for computational feasibility) to determine covariate splits. $n_{C_j}$ and $n_P$ are the number of observations in child node $j$ and the parent node, and $\hat{\theta}_{C_j}$ are the estimated ITEs in child node $j$.
4. For covariates $x$, define the training set samples that fall into same node as $x$ in tree $\mathcal{T}$ as $L_b(x)$. The weight for each observation in $b$ is then calculated as $\alpha_{bi}(x) = \frac{\mathbf{1}(\{X_i \in L_b(x)\})}{|L_b(x)|}$.

With $B$ trees generated, the weight for each observation is $\alpha_i(x) = \frac{1}{B} \sum_{b=1}^{B} \alpha_{bi}(x)$. $\alpha_i(x)$ may be seen as the frequency that the $i$th sample falls in same node as $x$ across all trees in the forest. Therefore, $\sum_{i=1}^{n} \alpha_i(x) = 1$. The ITEs are estimated as a solution to the minimization problem $\left( \hat{\theta}(x), \hat{v}(x) \right) \in \operatorname{argmin}_{\theta,v} \{ \| \sum_{i=1}^{n} \alpha_i(x) \psi_{\theta,v}(O_i) \|_2 \}$ where $\hat{\theta}(x)$ is the

ITE for covariates $x$, $\hat{v}(x)$ is an optional nuisance parameter, and $\psi_{\theta,v}(O_i)$ is a general scoring function for the observed pair $O_i = \{Y_i, Z_i\}$.

In summary, GRF serves an abstraction to common forest methods with minor technical changes to address instances of substantial noise. Bootstrap sampling and random covariate sampling are retained from standard RF and CF, however key attributes such as splitting criteria and bagging are further generalized.

## 2.2 Bayesian Additive Regression Trees

Bayesian additive regression trees (BART) is the oldest method we consider (Chipman, George, and McCulloch 2010). Like CART, BART was first developed for prediction via nonparametric recursive partitioning. However, BART is specified as an additive sum-of-trees model within a bayesian framework. The additive model specification estimates linear relationships more accurately than individual tree models and attenuates interactions that individual tree models tend to overemphasize. Several works have proposed methods using bayesian single-tree models (Chipman, George, and McCulloch 1998; Denison, Mallick, and Smith 1998; G. 2004; Wu, Tjelmeland, and West 2007), including a bayesian backfitting Markov Chain Monte Carlo (MCMC) procedure (Hastie and Tibshirani 2000). BART combines elements of these and has been demonstrated to be an effective tool for estimating ITEs (J. L. Hill 2011).

BART is modeled as

$$Y = f(z, x) + \epsilon = \sum_{j=1}^{m} g(z, x; T_j, M_j) + \epsilon$$

where $(T_j, M_j)$ is the binary tree $T_j$ and corresponding $b$ terminal node parameters $M_j = \{\mu_1, \ldots, \mu_b\}$ set and $\epsilon \sim N(0, \sigma^2)$. $g(z, x; T_j, M_j)$ is the parameter associated with the terminal node that contains the pair $(z, x)$ for tree $(T_j, M_j)$. Analogous to boosting in which multiple weak learners contribute to one strong learner, each tree in the model is limited to small contributions to $f$. This is accomplished through a regularization prior that limits the influence of individual trees, suggesting that very high dimensional interactions are unrealistic. The prior can be specified by three components: $p(T_j)$, $p(\mu_{ij}|T_j)$, and $p(\sigma)$. The $T_j$ prior favors trees with few terminal nodes to avoid overfitting, while the $\mu_{ij}|T_j$ and $\sigma$ priors shrink $M_j$ toward 0 and $\sigma$ to a value smaller than that estimated from least squares, respectively (J. L. Hill 2011). The purpose of these shrinkage priors is to assign strong probability to the region of the most plausible values for the parameter of interest without overconcentration or overdispersion (Chipman, George, and McCulloch 2010). With these specifications, the model is fit using a Bayesian backfitting MCMC algorithm that follows these steps:

1. Initialize the MCMC chain with $m$ single node trees (stumps) and set $\sigma^2 = 1$.
2. For each tree $j$ in 1,...,$m$:

- Calculate the partial residuals from the model fit without tree $j$, $R_j \equiv y - \sum_{k \neq j} g(x, z; T_k, M_k)$.
- Randomly perturb tree $j$ with one of grow, prune, change, or swap.
- Calculate $p(T_j|R_j, \sigma)$ and draw from the distribution to accept or reject the perturbation.
- Draw from $p(M_j|T_j, R_j, \sigma)$ to get the terminal node values for tree $j$.
3. Draw from the $\sigma$ prior conditional on all trees and the outcome.
4. Repeat steps 2-3 for a provided number of burn-ins and average the after burn-in samples over trees to get the final prediction for $(x, t)$, $\frac{1}{K}\sum_{k=1}^{K} f_k^*(x, t)$ where $f^*(\cdot) = \sum_{j=1}^{m} g(\cdot; T_j^*, M_j^*)$.

In each MCMC iteration, trees are randomly perturbed by one of four actions. The "grow" action assigns a covariate split to a terminal node at random, while the "prune" action removes the children of a random parent node. The "change" and "swap" actions alter the prediction of terminal nodes by either randomly replacing the splitting rule of an internal node or swapping the splitting rules of a random parent-child internal node pair (McJames et al. 2023). Rather than fitting new trees to partial residuals, the algorithm uses these perturbations and the defined priors to accept or reject the changes to each tree via the Metropolis-Hastings procedure (Chipman, George, and McCulloch 2010). Figure 2 illustrates this algorithm for $m$ trees and $K$ MCMC iterations.

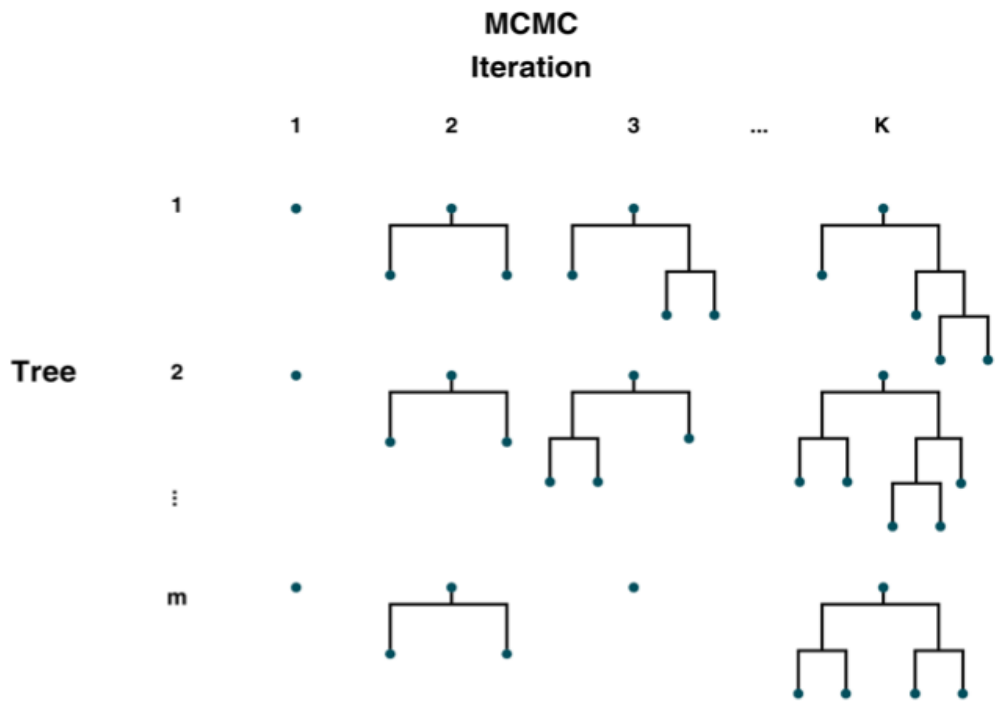

*Figure 2. Illustration of the BART Bayesian backfitting MCMC algorithm, inspired by Hastie and Tibshirani (2022).*

The key property of BART is its applicability as an effective "off the shelf" method. This is due to the default regularization prior specification and its anti-overfitting nature. Parameter tuning and cross-validation are often not necessary and consequently, the method is computationally less taxing and simpler to implement than others. Moreover, BART provides inference in the form of credible posterior intervals, which can be a more intuitive uncertainty metric than standard frequentist confidence intervals.

## 2.3 Bayesian Causal Forests

Bayesian causal forests (BCF) aims to improve upon BART in a similar sum-of-trees model (Hahn, Murray, and Carvalho 2020). The authors identified two issues with BART: "regularization-induced confounding" and high ITE estimate variability in the presence of homogeneity or moderate heterogeneity. Regularization-induced confounding occurs when confounders are regularized out of the model outcome surface due to their lack of predictive power (Hahn et al. 2016). This is an issue when the confounders serve the role of reducing bias in the exposure-outcome relationship. Hahn et al. propose including an estimate of the propensity score as a covariate to address this issue. High ITE estimate variability occurs when the heterogeneity signal is low with default BART and the authors propose reparameterizing the outcome surface into two independent BART ensembles as a fix.

The BCF model is expressed as

$$Y = f(z, x) + \epsilon = \mu(x, \hat{\pi}(x)) + \tau(x)z + \epsilon$$

where $\mu$ is the BART component that explains the relationship between $x$ and $Y$, and $\tau$ is the BART component that represents the treatment effect as a function of covariates $x$. This reparameterization allows for different covariate sets for $\mu$ and $\tau$ which is beneficial when it is known that set of moderators and confounders are not equivalent. The number of trees can also differ between $\mu$ and $\tau$, enabling users to customize the complexity of the components (McJames et al. 2023). Lastly, $\mu$ and $\tau$ have independent BART priors that can provide different regularization. The default BCF $\tau$ prior provides stronger regularization than the $\mu$ prior, favoring homogeneity unless there is strong evidence to the contrary.

The Bayesian backfitting MCMC algorithm for BCF is identical to that of BART, with slightly different priors and two different BART models run within each MCMC iteration. Overall, BCF provides adjustments for identified problems with standard BART and seeks to improve the accuracy of ITE estimation.

# 3 Implementation of the selected EMM ML methods

In this section, we review the currently available tools to implement these EMM ML methods in R version 4.2.2 (R Core Team 2023) and apply them to case study related to the effect of drought on stunting using the Demographic and Health Survey (DHS) data (see details below). All tunable parameters for each method were set to default and the code for these implementations can be found in the appendix.

## 3.1 Data

Table 2: *Descriptive statistics*

| | | Stunted Child Growth | |
|---|---|---|---|
| **Variable** | **Overall**, N = 345,499[1] | **Not Stunted**, N = 212190 (61%)[1] | **Stunted**, N = 133309 (39%)[1] |
| Exposure | | | |
| Drought | 50,310 (15%) | 30,034 (14%) | 20,276 (15%) |
| Covariates (Effect Modifiers/Confounders) | | | |
| Child Sex - Male | 174,324 (50%) | 103,099 (49%) | 71,225 (53%) |
| Child Age - Under 2 | 127,378 (37%) | 84,656 (40%) | 42,722 (32%) |
| Child Birth Size - Small | 62,218 (18%) | 34,449 (16%) | 27,769 (21%) |
| Child Breastfed - Never | 7,463 (2.2%) | 4,558 (2.1%) | 2,905 (2.2%) |
| Mother's Education - None | 174,528 (51%) | 99,828 (47%) | 74,700 (56%) |
| Mass Media Consumption - Yes | 156,872 (45%) | 105,636 (50%) | 51,236 (38%) |
| Single Mother - Yes | 27,128 (7.9%) | 16,348 (7.7%) | 10,780 (8.1%) |
| Agricultural Occupation - Yes | 161,724 (47%) | 87,741 (41%) | 73,983 (55%) |
| Residence - Rural | 249,258 (72%) | 142,971 (67%) | 106,287 (80%) |
| Wealth - Poor | 158,259 (46%) | 83,998 (40%) | 74,261 (56%) |

[1]n (%)

*All variables recorded as binary*

We use the DHS data set to demonstrate the application of these methods. These data describe health, disease, and wellness measures for over 90 low- and middle-income countries (LMIC). For our analysis, we restrict the data to observations collected between 2014-2019 of children under 3 years of age. Data from the Climate Hazards Group InfraRed Precipitation with Station (CHIRPS) database was merged with the DHS data in order to assess the impact of climate shocks on the health status of LMIC children under 3.

The outcome in our analysis is stunted child growth and the exposure is drought. The covariates are child sex, age, birth size, breastfed status, the mother's education level, single status, occupation, and family media consumption, rural residence, and wealth level. While we choose to focus on this particular set of covariates, the methods and corresponding interpretation are generalizable to higher dimensions of EMM. Table 2 lists the variables used in this analysis and their prevalence in the data. All variables are coded as binary indicators and missingness has been removed such that all observations are complete.

## 3.2 ITE Estimators

### 3.2.1 Generalized Random Forests

GRF may be implemented with the *causal_forest* function in the R package *grf* (Tibshirani et al. 2023). The required inputs for the function are the matrix of covariates, the outcome vector, and the exposure vector. Tuneable parameters include the number of trees to be grown in the forest, minimum node size and depth, and honest splitting ratio. After the forest has been generated, the *predict* function is used to obtain the output vector of ITEs expressed as risk differences and *average_treatment_effect* can be used to estimate the ATE. Additionally, there are several commonly used tools for GRF to assess EMM. A variable importance metric for each covariate is provided by the function *variable_importance*, representing a weighted sum of the number of times each covariate was used in a tree node split. In practice, this metric can be used to readily identify potential effect modifiers, as a large variable importance value indicates large influence in the ITE estimation. *test_calibration* conducts a "best linear predictor" (BLP) analysis that serves as an omnibus calibration test of the quality of the ITE estimates and presence of heterogeneity within the data (Athey and Wager 2019; Chernozhukov et al. 2018). This is done by regressing $Y_i - \hat{m}^{(-i)}(X_i)$ against $C_i$ and $D_i$ where $C_i = \bar{\tau}\left(T_i - \hat{e}^{(-i)}(X_i)\right)$ and $D_i = \left(\hat{\tau}^{(-i)}(X_i) - \bar{\tau}\right)\left(T_i - \hat{e}^{(-i)}(X_i)\right)$. The notation "$(-i)$" represents the out-of-bag estimate for unit $i$, $\hat{m}^{(-i)}(x)$ and $\hat{e}^{(-i)}(x)$ are the estimated out-of-bag mean outcome and propensity scores for $x$ from separate regression forests, $\hat{\tau}^{(-i)}(x)$ is the out-of-bag ITE estimate for $x$, and $\bar{\tau}$ is the average out-of-bag ITE estimate. The coefficient of $C_i$ can be interpreted as the quality of the ATE estimate while the coefficient of $D_i$ represents the quality of the ITE estimates. When the coefficient of $C_i$ or $D_i$ is close to 1 in magnitude, the ATE or ITEs are well calibrated, respectfully. Moreover, if the coefficients of $C_i$ or $D_i$ are positive and significant, then there is evidence to reject the null hypothesis of no main effect or EMM in the data. A second stage regression analysis can also be performed with the function *best_linear_projection* that regresses the ITE estimates on prespecified effect modifiers. The resulting model coefficients are doubly robust CATEs for subgroups defined by the provided effect modifiers. Lastly, a technique known as "fit-the-fit" can be performed in which the estimated ITEs are used as the outcome in a CART model. This provides a simple visualization to determine which covariates contribute most to the variability in the ITEs. Example applications of these techniques can be seen in papers such as Shiba et al. (2021), Y. Zhang, Lia, and Ren (2022), M. Wang and Yang (2022), and Goldman-Mellor et al. (2022).

*Table 3: Generalized Random Forests results*

| Estimand | Estimate (p-value or 95% CI) |
|---|---|
| **Best Linear Predictor Calibration** | |
| Mean forest prediction | 0.988 (<0.001) |
| Differential forest prediction | 0.385 (<0.001) |
| **Second Stage Regression** | |
| Mother's Education - None | 0.042 (0.033, 0.051) |
| Residence - Rural | 0.019 (0.009, 0.029) |
| **Variable Importance** | |
| Mother's Education | 61.9% |
| Residence | 11.1% |
| Mass Media Consumption | 6% |
| Single Mother | 4.5% |
| Agricultural Occupation | 3.9% |
| Child Age | 3.6% |
| Child Birth Size | 3.6% |
| Wealth | 3.2% |
| Child Sex | 2.2% |
| Child Breastfed | 0% |

Table 3 shows the results from BLP and second stage regression analyses, and lists the variable importance for each covariate. The mean forest prediction coefficient is close to 1 in magnitude and statistically significant, indicating that there is an overall effect of drought on stunted child growth measured by the ATE and the GRF captures this effect well. The coefficient for the differential forest prediction is statistically significant but not close to 1 in magnitude. This implies that heterogeneity exists in the data but it was not accurately captured by the GRF ITEs. For the purpose of demonstration, we perform the second stage regression on the subgroups of the covariates with over 10% variable importance. We see that maternal education and residence status are the most important covariates in the GRF tree node splits. The second stage regression finds that the CATEs for those whose mothers have no education and those who live in a rural area are 4.2% (95% CI: [3.3%, 5.1%]) and 1.9% (95% CI: [0.9%, 2.9%]), respectively.

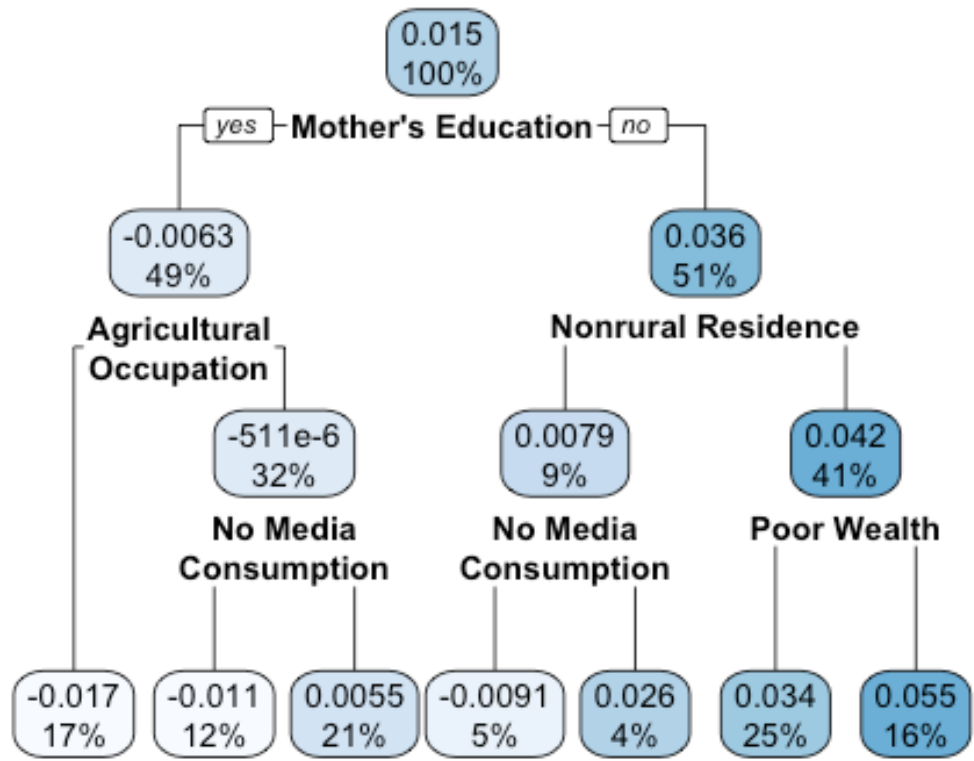

*Figure 3. Generalized Random Forests CART (classification and regression tree)*

Figure 3 shows the CART of GRF ITEs with a set maximum node depth of 3. The ITEs are first split on maternal education. 49% of the observations have some maternal education and an average ITE of -0.63% while 51% of the observations have no maternal education and an average ITE of 3.6%. Rural residence is also represented in a level 1 node split with child node averages of 4.2% and 0.79%.

### 3.2.2 Bayesian Additive Regression Trees

Several packages in R exist to implement BART. The most prominent include *dbarts* (Dorie, Chipman, and McCulloch 2023; Dorie 2023) and its corresponding causal inference derivative *bartCause* (J. L. Hill 2011), and *BART* (Sparapani, Spanbauer, and McCulloch 2021). These packages are modern iterations of *BayesTree* (H. Chipman and McCulloch 2016) and *bartMachine* (Kapelner and Bleich 2016). We choose *BART* for our application, but any of these packages may be used to implement BART. The function *wbart* generates the model for continuous outcomes while *lbart* and *pbart* are used for dichotomous outcomes (logistic and probit links, respectively). To estimate ITEs, 2 counterfactual data sets are created where the exposure is present and absent for all observations. *w/l/pbart* can then be run on each data set, representing estimated ITEs for the exposed and unexposed pseudo-populations. Like those of GRF, BART ITEs are expressed as risk differences by default. The required inputs to the function are the matrix of covariates and outcome vector to train the model, and the matrix of counterfactual covariates to estimate the ITEs. The desired output vector of ITEs can be obtained by averaging the estimated outcome across MCMC iterations and taking the difference between the exposed and unexposed ITE sets. For dichotomous or other outcomes in which the link function is not the identity, ITEs may be transformed back to the additive scale to estimate absolute risk (for a binary outcome, we use the inverse logit function $expit(x) = 1/(1 + exp(-x))$). Tuneable parameters of note are the number of burn-in MCMC iterations as the threshold for posterior convergence and the number of MCMC iterations to save after burn-in. The ITEs may be analyzed using a fit-the-fit CART approach, and several variable importance metrics and selection methods have been proposed to determine covariate influence in BART models (Bleich et al. 2014; Luo and Daniels 2021; Inglis, Parnell, and Hurley 2021; Carvalho et al. 2023). Example applications of BART for EMM analysis can be found in Blette et al. (2023), Carnegie, Dorie, and Hill (2019), L. Hu et al. (2021), and Kraamwinkel et al. (2019).

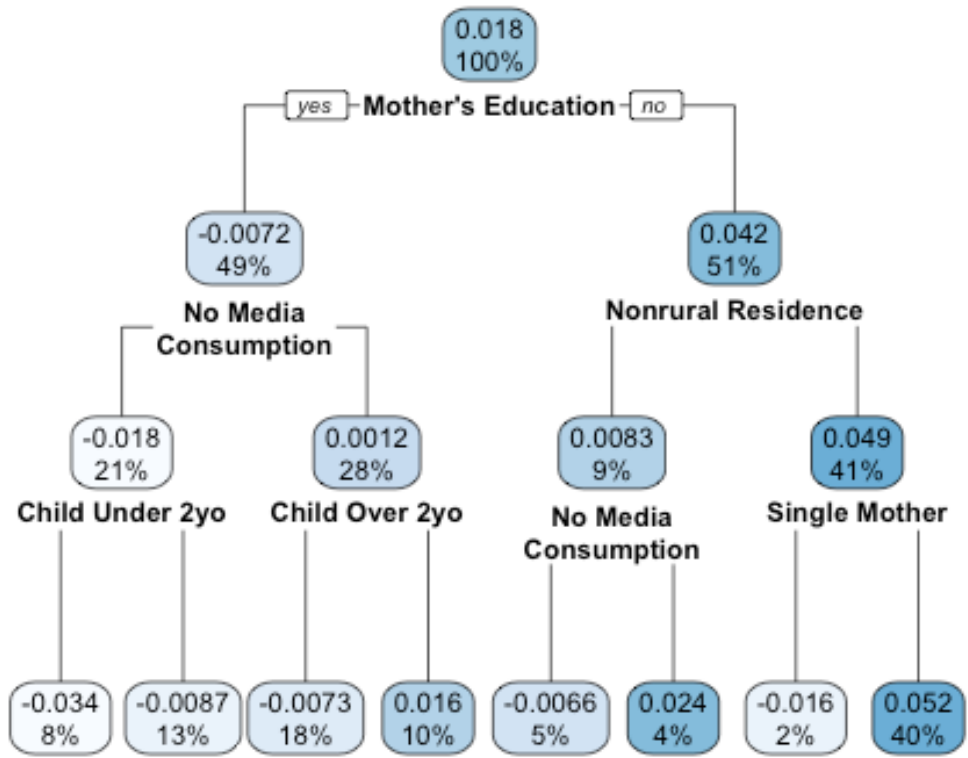

Figure 4. BART CART

The BART fit-the-fit CART is shown in figure 4. Maternal education is again the first covariate used to split the ITEs and the average child node ITEs are similar to those from GRF. Rural residence features as level 1 splitting variable and the child node ITEs agree with the GRF results.

### 3.2.3 Bayesian Causal Forests

BCF is implemented with the package and identically named function *bcf* (Hahn, Murray, and Carvalho 2020). The required inputs are the matrix of covariates, outcome vector, exposure vector, and estimated propensity score. We choose to estimate the propensity score with a logistic regression model where the exposure is modeled as a function of the additive effects of the covariates for simplicity. Like BART, the burn-in MCMC iteration threshold and the number of saved MCMC iterations after burn-in are tuneable parameters, and the risk difference ITE estimates are obtained by averaging the estimated outcome over MCMC iterations. The primary tool to analyze EMM using BCF is the fit-the-fit CART approach. Currently, we are not aware of other example applications of BCF for EMM analysis.

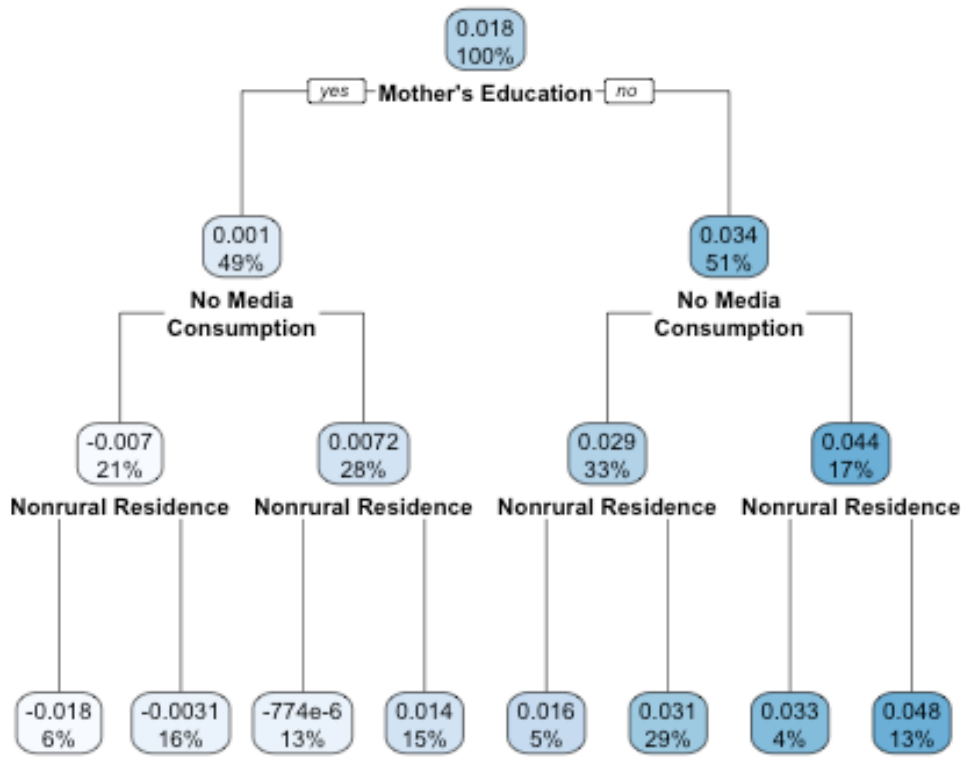

*Figure 5. BCF CART*

Figure 5 shows the BCF CART. The covariate used in the first split rule is maternal education and the average child node ITEs are similar to those from the GRF and BCF CARTs. Unlike the GRF and BCF CARTs, rural residence is not used as a level 1 splitting variable. However, it is used as a level 2 splitting variable for all nodes in the tree.

### 3.2.4 Comparison of ITE estimator methods

The most common technique for EMM analyses using these methods is the fit-the-fit CART modeling of the estimated ITEs. The CART model serves as a simple, automated tool to identify subgroups where the ITE estimates most differ. Most applications of this technique use a maximum node depth of 3 for interpretability (J. Hill and Su 2013). Variable importance metrics can also be used to identify potential effect modifiers and second stage regression allows for CATE estimation of specified subgroups. It is also common to plot the distribution of ITEs for different subgroups to visualize potential heterogeneity. Below are plots of each method's ITE distribution for different subgroups of potential effect modifiers. For all three methods, there is a visible increased risk for units with no maternal education compared to those who had education, and for those whose residence status is rural compared to non-rural.

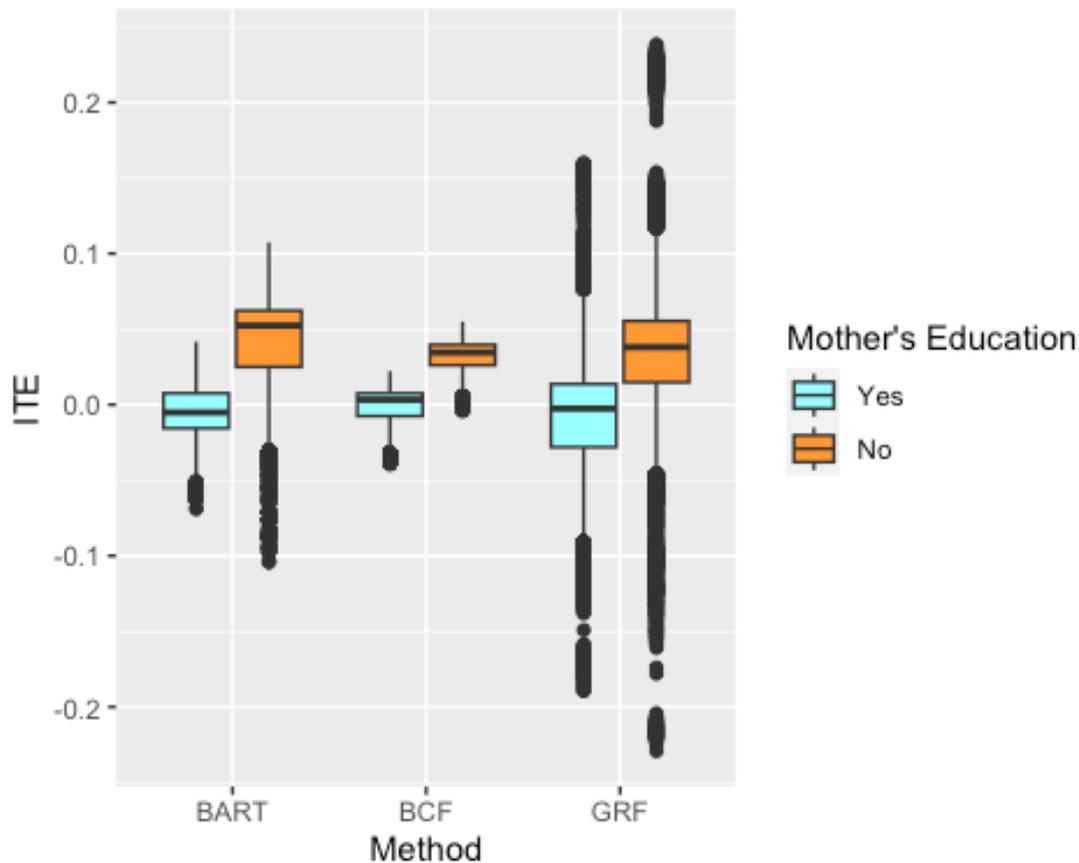

*Figure 6. ITE comparison for GRF/BART/BCF across levels of mother's education*

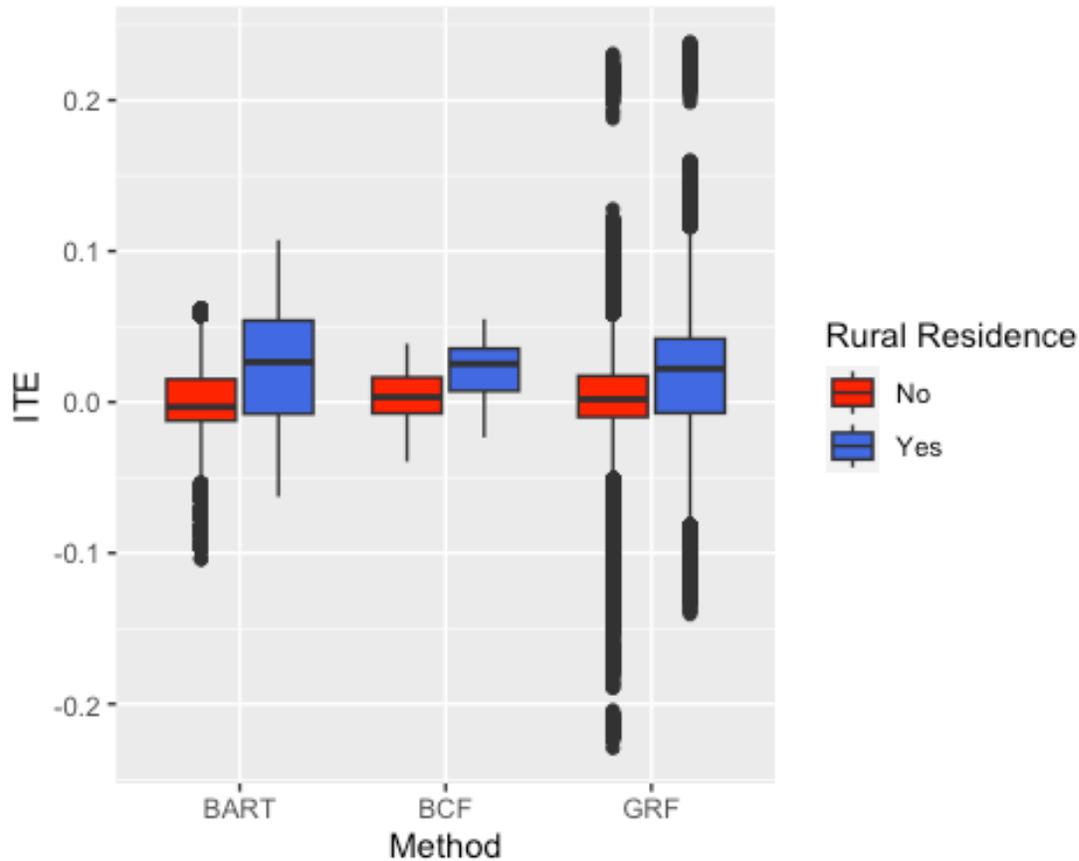

*Figure 7. ITE comparison for GRF/BART/BCF across levels of residence status*

### 3.3 Comparison with traditional methods

As discussed in section 1, there are several traditional methods for EMM analysis. We choose to estimate CATEs with a stratified analysis of logistic regression models adjusted for all covariates and the CATEs are represented as odds ratios. To compare results to those from the ML methods, we use the subgroups defined by the two levels of maternal education. We also run Cochran's chi-squared test to determine if there is evidence to reject the null hypothesis of no heterogeneity.

Table 4: *Traditional methods results*

| Estimand | Full Sample | Mother's Education - None | Mother's Education - Some |
|---|---|---|---|
| Risk Difference (95% CI) | 0.02 (0.015, 0.025) | 0.034 (0.028, 0.041) | -0.007 (-0.014, 0) |
| Risk Ratio (95% CI) | 1.053 (1.04, 1.065) | 1.081 (1.066, 1.096) | 0.979 (0.96, 0.999) |
| (C)ATE (95% CI)[a,b] | 1.066 (1.045, 1.087) | 1.151 (1.121, 1.182) | 0.97 (0.941, 1) |
| Cochran's Q test statistic (p-value) | | 65.492 (< 0.001) | 65.492 (< 0.001) |

[a]Expressed as odds ratio
[b]Adjusted for covariates

Table 4 gives the estimands and corresponding 95% confidence intervals for the traditional analysis methods and results from Cochran's chi-squared test for heterogeneity. The first two rows give the unadjusted risk difference and relative risk estimates in the full data, no maternal education subgroup, and some maternal education subgroup. We find that there is a significant increased risk of stunted child growth due to drought in the full sample (RD: 2% [1.5%, 2.5%]; RR: 1.05 [1.04, 1.07]). The risk is larger in the no maternal education subgroup (RD: 3.4% [2.8%, 4.1%]; RR: 1.08 [1.07, 1.1]), and there is no strong evidence of an increased risk in the some maternal subgroup (RD: -0.7% [-1.4%, 0%]; RR: 0.98 [0.96, 1]). The third row gives the ATE in the full sample and CATEs from the maternal education subgroups. These estimates also suggest a significant increased risk of stunted child growth due to drought in the full sample (OR: 1.07 [1.05, 1.09]) and a larger risk in the no maternal education subgroup (OR: 1.15 [1.12, 1.18]). Again we find no strong evidence of an effect of drought on stunted child growth in the some maternal education subgroup (OR: 0.97 [0.94, 1]). Lastly, Cochran's chi-squared test suggests there is evidence to reject the null hypothesis of no EMM between the maternal education subgroups (p < 0.001). These findings agree with those found by the ML methods and strengthen the case for maternal education as an effect modifier.

## 4 Discussion

In this overview, we summarized three recently developed ML methods for EMM analysis that have been used in various quantitative social sciences disciplines but to a lesser extent in epidemiology data. We also discussed current implementation in R and applied them using a case study focusing on the effect of drought on stunting among children in multiple Sub-Saharan countries. These nonparametric, data-driven methods allow for flexible modeling of nonlinear outcome surfaces and high-dimensional interactions that are impractical to test by hand. At the time of this overview, generalized random forests and bayesian additive regression trees are the most commonly used ML methods for EMM analyses, while bayesian causal forests is a novel derivative. Each method primarily estimates ITEs expressed as risk differences and as a consequence, requires additional analysis tools to identify interpretable heterogeneous subgroups and estimate CATEs.

It is important to mention that no algorithm can automatically select what constitutes a true effect modifier which needs to be based on the pre-existing knowledge regarding a specific exposure-outcome relationship. While the concepts of confounding and effect modification are fundamentally different, it is known that confounders (that are minimally associated with the outcome of interest, without being a collider variable) are supposed to be effect modifiers on at least one scale (additive or multiplicative) (Lash et al. 2020). Therefore, we suggest it is reasonable to consider all confounders (and their multiple combinations and functional forms) as potential effect modifiers when exploring high-dimensional EMM analyses.

"Fit-the-fit" CART and variable importance are tools that are often used to efficiently identify potential effect modifiers. However, it is also important to note that these techniques do not guarantee that identified covariates are true effect modifiers. Covariates may have large influence in fit-the-fit CART models or large variable importance due to high correlation with true effect modifiers (Jawadekar et al. 2023). Variable importance should not be interpreted as the proportional influence of heterogeneity or the likelihood of a covariate being a true effect modifier. Rather, these tools are helpful for identifying covariates for which variability in the ITE estimates is high with no prior knowledge of true effect modifiers. Second stage regression and plotting ITE densities across covariate groups are also used to assess EMM. However, these techniques require manual specification of potential effect modifiers.

We have not discussed other methods that may be used to identify heterogeneous subgroups and estimate CATEs due to their novelty and lack of use in empirical settings. However, we would like to mention some recently developed alternative approaches. In particular, the causal rule ensemble (CRE) (Lee, Bargagli-Stoffi, and Dominici 2021) and multilevel analysis of individual heterogeneity and discriminatory accuracy (MAIHDA) (Rodriguez-Lopez et al. 2023) are methods that can automate effect modifier identification. The CRE accomplishes this through a procedure of estimating ITEs, generating heterogeneous subgroups from the ITE estimates, and selecting the most important heterogeneous subgroups through penalized regression. Several ML methods are used in this ensemble procedure such as random forests and LASSO, and any of GRF, BART, or BCF may be used during the ITE estimation step. MAIHDA implements a mixed-effects regression model where subgroups are treated as random-intercepts. Heterogeneous subgroups can be identified by the magnitude of the estimated random effects and CATEs are estimated as linear combinations of fixed and random effects. These methods are not commonly used for EMM analyses yet, but are promising methods for automating heterogeneous subgroup identification.

There are several limitations of this overview that motivate future work. We chose to demonstrate the application of these methods using observational data. However, post-hoc heterogeneity analyses of clinical trial data are often underpowered because the data are collected to power the main treatment effect. Extending these methods outside of the observational setting to clinical trial data with techniques such as data fusion and integration remains a methodological challenge (L. Zhang et al. 2018 ). Moreover, we do not evaluate the performance of the methods and we refer to other work comparing methods that do so (Wendling et al. 2018; Dorie et al. 2019; McConnell and Lindner 2019; L. Hu et al. 2021; Liu 2022; A. Hu 2023). Ultimately, all of the considered methods are useful tools for exploring heterogeneity within real world data. The data-driven nature of these methods distinguishes them as helpful tools for initial exploration of effect modification, while traditional methods better serve as tools for confirmatory analyses.

In conclusion, machine learning for effect measure modification is a burgeoning and promising field of study. As such, there is a constant need for simple interpretation of newly developed tools to increase their accessibility for applied researchers. This overview provides this interpretation and guides readers on implementing these tools and other supplemental analysis techniques in their own research. We hope that it serves as a useful reference for researchers in public health and adjacent disciplines.

# Appendix

## A.1 GRF application

```r
# set.seed(1102)

# Run GRF (grf::causal_forest)
grf_results <- causal_forest(X = X,
                             Y = y,
                             W = z)

# Get ITEs
grf_ite <- predict(grf_results, estimate.variance = T)

# GRF BLP calibration
test_calibration(grf_results)

# GRF variable importance
(grf_varimp <- tibble(variable = names(X),
                      importance = variable_importance(grf_results)))

# GRF CATEs
best_linear_projection(grf_results, A = dplyr::select(data, education_none))
best_linear_projection(grf_results, A = dplyr::select(data, rural_residence))
```

## A.2 BART application

```r
set.seed(1102)

# Make counterfactuals for BART algorithm to test
data1 <- data %>%
    mutate(drought = 1)

data0 <- data %>%
    mutate(drought = 0)

# Run BART (BART::lbart)
## Run for exposed
bart_results1 <- lbart(x.train = as.data.frame(dplyr::select(data, -outcome))
,
                       y.train = dplyr::pull(data, outcome),
                       x.test = as.data.frame(dplyr::select(data1, -outcome))
)

## Run for unexposed
bart_results0 <- lbart(x.train = as.data.frame(dplyr::select(data, -outcome))
,
                       y.train = dplyr::pull(data, outcome),
                       x.test = as.data.frame(dplyr::select(data0, -outcome))
)
```

```r
bart_results1$yhat.train.collapse <- apply(bart_results1$yhat.train, 2, rbind
)
bart_results1$yhat.test.collapse <- apply(bart_results1$yhat.test, 2, rbind)
bart_results0$yhat.train.collapse <- apply(bart_results0$yhat.train, 2, rbind
)
bart_results0$yhat.test.collapse <- apply(bart_results0$yhat.test, 2, rbind)

# ITE estimates
bart_ite <- exp(colMeans(bart_results1$yhat.test.collapse)) /
    (1 + exp(colMeans(bart_results1$yhat.test.collapse))) -
    exp(colMeans(bart_results0$yhat.test.collapse)) /
    (1 + exp(colMeans(bart_results0$yhat.test.collapse)))
```

### A.3 BCF application

```r
set.seed(1102)

# Estimate propensity score for BCF
ps_mod_formula <- as.formula(paste(exposure, paste(names(X), collapse = " + "
), sep = "~"))
ps_mod <- glm(ps_mod_formula,
              family = binomial,
              data = data)
logit_ps <- predict(ps_mod, newdata = data, type = "response")

# Format design matrix for BCF
X_BCF <- makeModelMatrixFromDataFrame(X)

# Run BCF (bcf::bcf)
bcf_results <- bcf(y = y,
                   z = z,
                   x_control = X_BCF,
                   pihat = est_ps,
                   nburn = 500,
                   nsim = 500)

# ITE estimates
bcf_ite <- colMeans(bcf_results$tau)
```